\documentclass[aps,12pt,twoside,a4paper]{revtex4}
\usepackage{epsfig}
\usepackage{graphicx}
\usepackage{amsmath,amssymb,color}
\usepackage{footnote}
\usepackage[english]{babel}

\parskip=\medskipamount

\newcommand{\eq}[1]{(Eq. \ref{#1})}
\newcommand{\fig}[1]{Fig.\ref{#1}}
\newcommand{\be}{\begin{equation}}
\newcommand{\ee}{\end{equation}}

\newcommand\disp{\displaystyle}

\newcommand{\la}{\left<}
\newcommand{\ra}{\right>}

\begin{document}

\title{Fractal dimension meets topology: statistical and topological properties of globular macromolecules with volume interactions}

\author{Alexey M. Astakhov}
\email{astahov.aleksey@physics.msu.ru}
\affiliation{Physics Department of the Lomonosov Moscow State University, Moscow, Russia}
\affiliation{N.N. Semenov Institute of Chemical Physics RAS, Moscow, Russia}
\author{Vladik A. Avetisov}
\affiliation{N.N. Semenov Institute of Chemical Physics RAS, Moscow, Russia}
\author{Sergei K. Nechaev}
\affiliation{Interdisciplinary Scientific Center Poncelet (CNRS UMI 2615), Moscow, Russia}
\affiliation{P.N. Lebedev Physical Institute RAS, Moscow, Russia}
\author{Kirill E. Polovnikov}
\affiliation{Institute for Medical Engineering and Science, Massachusetts Institute of Technology, Cambridge, MA 02139}
\affiliation{Skolkovo Institute of Science and Technology, 143026 Skolkovo, Russia}

\begin{abstract}

In the paper we investigate statistical and topological properties of fractional Brownian polymer chains, equipped with the short-range volume interactions. The attention is paid to statistical properties of collapsed conformations with the fractal dimension $D_f\ge 2$ in the three-dimensional space, which are analyzed both numerically and \textit{via} the mean-field Flory approach. Our study is motivated by an attempt to mimic the conformational statistics of collapsed unknotted polymer rings, which are known to form compact hierarchical crumpled globules (CG) with $D_f=3$ at large scales. Replacing the topologically-stabilized CG state by a self-avoiding fractal path adjusted to the fractal dimension $D_f=3$ we tremendously simplify the problem of generating compact self-avoiding conformations since we wash out the topological constraints from the consideration. We make use of the Monte-Carlo simulations to prepare the equilibrium ensemble of swollen chains with various fractal dimensions. A combination of the Flory arguments with statistical analysis of the conformations from simulations allows one to infer the dependence of the critical exponent of the swollen chains on the fractal dimension of the seed chain. We show that with the increase of $D_f$, typical conformations become more territorial and less knotted. Distributions of the knot complexity, $P(\chi)$ for various fractal dimensions of the swollen chains suggest a close relationship between statistical and topological properties of fractal paths with volume interactions.

\end{abstract}

\maketitle

\section{Introduction}

\subsection{From topology to hierarchy: Reminder of a crumpled globule concept}

The crumpled (fractal) globular state of a macromolecule resembling statistical Peano or Hilbert curves \cite{Peano} with the fractal dimension $D_f = 3$ in the three-dimensional space, was proposed in \cite{GrosbergNechaevShakh} as an equilibrium structure of unknotted polymer ring in a poor solvent. This state is notably different from the classical Lifshitz globule (formed by a linear chain without topological constraints), which is not self-similar and is highly knotted. Nonlocal nature of topological interactions overwhelmingly complicates analytical treatment of the crumpled polymer states. In particular, the interplay between topological and fractal properties of crumpled conformations remains to be poorly understood.

The condition for a polymer ring to be globally unknotted induces hierarchical collapse of a ring into a set of self-similar crumples in a confined volume or under poor solvent conditions. Basic ideas behind this process are as follows. Consider a closed unknotted non-self-intersecting long polymer chain. In a poor solvent (below the $\theta$-point), there exists a certain critical chain length, $g^*$, depending on the temperature and the energy of volume interactions, such that chains with lengths larger then $g^*$, collapse. As it was shown in \cite{GrosbergNechaevShakh}, for sufficiently long chain, units of length $\sim g^*$ may be regarded as new "block monomers" (crumples of the minimal scale). Consider now a part of a chain containing several block monomers. Such new part of the chain should collapse "in itself" -- that is, it should form a crumple of the next scale if other parts of the chain do not belong to a given part. A chain of new sub-blocks (crumples of the new scale) collapses again and again "in itself" until the chain as a whole forms a crumple of the largest scale containing all initial monomers. It has been shown in \cite{AstakhovNechaevPolovnikov} that stochastic formation of irreversible links during the collapse prohibits intermixing of the chain segments of different scales and facilitate stabilization of the hierarchy of crumples. The line representing the chain trajectory is a 3D statistical analog of the well-known self-similar hierarchical Peano curve. 

The interest of biologists to the crumpled structures was heated by the work \cite{GrosbergRabinHavlin} in which the authors conjectured that eukaryotic chromosomes, at least at some scales, might benefit from being unentangled. The hypothesis proposed in \cite{GrosbergRabinHavlin} suggested an answer to one of the principal issues in cell biology: how can long DNA chain reversibly collapse and decollapse in a small volume of the cell nucleus not being hopelessly tangled. The attitude towards the crumpled globule hypothesis has been substantially increased with the publication of the paper \cite{Mirny2009}, in which the average probability of spatial contacts between fragments of chromatin has been experimentally evaluated in the genome-wide chromosome conformation capture experiment (so-called "Hi-C" method) \cite{Dekker}. A mean-field argument suggests that a self-similar chain with the fractal dimension $D_f$ in the tree-dimensional space is characterized by the contact probability $P(s) \sim s^{-3/D_f}$ (see \cite{Mirny2009, MirnyImakaevNech} for the details), where $s$ is the distance along the chain contour, and the crumpled globule yields the scaling $P(s) \sim s^{-\alpha}$ with $\alpha=1$. Clearly, such a slow large-scale decay is impossible to realize in a system with finite excluded volume due to logarithmic divergence of the total number of contacts per monomer. In computer simulations, however, larger exponents have been reported, $\alpha \approx 1.1$ -- see \cite{Halverson11, RE14,RE19,Ge} for simulations of equilibrium ring chains. Notably, in Hi-C experiments \cite{Mirny2009, Halverson14} the observed values of $\alpha$ are typically close to unity (being slightly above $\alpha=1$), providing a strong evidence in favour of the crumpled globule hypothesis. Furthermore, several recent analyses of experimental data revealed that interphase chromosomes are largely unentangled and unknotted \cite{Tavares2020,Dimos}.

Despite the hierarchical structure of the crumpled globule actually reproduces the geometry of the Peano (or Hilbert) curve, the contact probability of the latter is $P_H(s) \sim s^{-4/3}$, in sharp contrast with the scaling of the crumpled globule \cite{Smrek13,Smrek15}. This contradiction points out on the affiliation of these exponents to different universality classes. The reduced contact probability of the Hilbert curve indicates that its segments are much less interdigitating than the crumples of the unknotted ring. The interpenetration of crumples is associated with the smoothness of their boundaries: in case of rings the boundaries turn out to be very developed with corresponding surface fractal dimension $d_b \approx 3$ (the amount of surface monomers, $N_b(s)$, of a subchain scales with its spatial size, $r(s)$, as $N_b(s) \sim r(s)^{d_b}$), while domains of the Hilbert curve are smooth, providing $d_b=2$. The works \cite{Halverson11, Smrek13} establish a simple scaling relation between the fractal dimension of the surface and the exponent of contact probability, $d_b/3+\alpha=2$, elucidating the underlying distinction between different space-filling curves.

Fractal folding immediately opens up an opportunity to mimic the ensemble of the crumpled trajectories by the ensemble of the 3D fractional Brownian motion (fBm) paths, which is the only process that is self-similar, stationary and Gaussian. Let us recall that fBm, $B_H(t)$, according to B. Mandelbrot \cite{mandelbrot}, is an average of increments of ordinary Brownian motion weighted with a certain non-local (algebraic) memory kernel. It can be defined by the Weyl integral as follows:
\be
B_H(t_2) - B_H(t_1) = \frac{1}{\Gamma(H+1/2)}\left[\int_{-\infty}^{t_2} (t_2-s)^{H-1/2}dB(s) -
\int_{-\infty}^{t_1} (t_1-s)^{H-1/2}dB(s) \right]
\label{fract_def}
\ee
where $H$ is the Hurst parameter, $0 < H < 1$. It is easy to see that the value $H=\frac{1}{2}$ brings us back to the ordinary Brownian motion. Importantly, a fBm is an example of a Gaussian process with a long-ranged memory, provided by non-local weights in \eq{fract_def}. Compared to the standard Brownian motion, it is non-Markovian, since the increments of fBm are scale-free variables. Moreover, fBm is an example of a Gaussian process, that violates strong mixing condition.

Space-filling fBm paths with the Hurst exponent $H=1/3$ are characterized by the contact probability exponent $\alpha=1$, thus, they could approximate surface roughness of unknotted polymer rings. As it was shown in \cite{fbm1}, it is possible to construct a pairwise Hamiltonian to describe self-similar polymer chains with arbitrary fractal dimension $D_f \le 2$. The resulting chains are Gaussian and share statistical properties with the fBm trajectories. In the fBm polymer chains the monomers interact \textit{via} harmonic pairwise couplings, which have to be specially designed to stabilize the large-scale fractal folding. Analytic tractability of the fBm model allows one to understand dynamical properties of crumpled polymers, as well as the implications of the long-ranged memory on many-body contacts in such chains \cite{polovnikov19_pre}. Subdiffusive fractional Brownian motion with $H<1/2$ turns out to be a very prolific model of chromatin dynamics \cite{polovnikov18,tammPolovnikov,amitai13}. However, fBm conformations are phantom and Gaussian, being in sharp contrast with properties of equilibrium unknotted rings \cite{RE19}. To this end, the study of \textit{swollen} fBm chains, equipped with short-range volume interactions, is highly demanded.

The connection between the non-local action of topological constraints and the space-filling folding remains rather obscure. We understand how the trivial topology of collapsed rings leads to folding with the fractal dimension $D_f=3$ in a 3D space. However the reverse question: "Whether globular polymers with $D_f>2$ in 3D are less knotted than the equilibrium globules?" -- is still open. In this paper, on the basis of the fBm Hamiltonian model proposed in \cite{fbm1}, we construct a new class of fractal conformations with soft volume interactions and study it by means of the Flory theory and Monte-Carlo simulations. In the next Section we elucidate what follows from the cross-fertilization of topology and fractal folding without and with volume interactions.

\subsection{Fractional Brownian statistics of chains with the effective pairwise Hamiltonian}

It has been mentioned already that interactions in topologically stabilized globular polymers such as collapsed rings, are substantially nonlocal. The very fact that topology cannot be screened drastically complicates the construction of the explicit microscopic Hamiltonian for the rings and to-day the development of analytic theory of topologically interacting polymers from the first principles remains to be a challenging problem.

One of the possible ways to take into account topological interactions in the globular phase is to link the chain topology to the fractal dimension. Namely, the absence of knots in a crumpled globule can be considered as a "sufficient condition" to have $D_f=3$ in the 3D space. However, we might think of the "necessary condition", considering formation of knots on closed dense curves with $D_f=3$. If the loops with $D_f=3$ are statistically less knotted than the ones with $D_f<3$, it would be possible to describe the trivial (or almost trivial) knot topology of globular polymers by fixing the fractal dimension, $D_f$, leaving behind the microscopic description of topological interactions. Such a tremendous simplification might be useful in many problems where topological interactions play a crucial role, however their exact account is a very complex technical problem -- for example, in the theory of the coil--crumpled globule phase transition.

In \cite{fbm1} it has been derived a pairwise Hamiltonian for Gaussian polymer chains with arbitrary fractal dimension $2\le D_f\le 3$. The Hamiltonian implies pairwise quadratic interactions $V({\bf r}_k,{\bf r}_m)$ acting between monomers $k$ and $m$ of the chain,
\be
V({\bf r}_k, {\bf r}_m) = a_{km} \left({\bf r}_k-{\bf r}_m\right)^2
\label{eq:01}
\ee
The large-scale statistics of chain conformations in the collective potential $V=\sum_{k,m}V({\bf r}_k, {\bf r}_m)$ is identical to the statistics of the fractional Brownian motion (fBm) \cite{mandelbrot}. In particular, the potential $V$ at equilibrium produces fBm-like distribution of the monomer-to-monomer distance
\be
P({\bf r}_k, {\bf r}_m) = \left(\frac{3}{2\pi b^2 |k-m|^{2/D_f}}\right)^{3/2} \exp\left(-\frac{3 ({\bf r}_k - {\bf r}_m)^2}{2 b^2 |k-m|^{2/D_f}}\right)
\label{eq:01distr}
\ee
with the mean-square distance
\be
\la(\mathbf{r}_k - \mathbf{r}_m)^2\ra \sim b^2 |k-m|^{2/D_f} \equiv b^2 s^{2/D_f}
\label{eq:01a}
\ee
where $s=|m-k|$ is the chemical distance between monomers, $b$ is the mean-squared size of one bead and $D_f$ is the fractal dimension lying in the interval $2 \le D_f \le 3$. In order to reproduce the behavior \eq{eq:01a}, the coupling constants $a_{km}(s)$ should decay asymptotically at $s\gg 1$ as
\be
a_{s} = c\, s^{-\gamma}; \quad \gamma > 2
\label{eq:02}
\ee
with $c>0$. The resulting large-scale fractal dimension of conformations is related to the exponent $\gamma$ by
\be
D_f = \left\{ \begin{array}{cl}
\disp \frac{2}{\gamma-2}, & \quad 2< \gamma \le 3 \medskip \\
\disp 2, & \quad \gamma > 3
\end{array} \right.
\label{eq:03}
\ee
The potential $V({\bf r}_k,{\bf r}_m)$ can be interpreted as the harmonic pairwise coupling with the spring rigidity $a_{km}$. If rigidities decay faster than $|m-k|^{-3}$, the large-scale statistics of the chain remains ideal despite the non-local interactions are present, which is equivocal to the central limit theorem in the Gnedenko-Kolmogorov formulation. Quadratic Hamiltonian with the potential \eq{eq:01} produces Gaussian phantom conformations with any fractal dimension $D_f\ge 2$ and the scale-free memory. The Gaussian property allows for a rigorous analytical treatment of the chain dynamics in complex media \cite{polovnikov18}, as well as of the statistics of many-body interactions \cite{polovnikov19_pre}.

\section{Model and simulation}

Fractional Brownian paths \cite{mandelbrot} share ambiguity of topological properties with the ordinary random walk: phantomness of trajectories does not allow to define a topological state of a particular path realization. A standard approach to the problem consists in introducing short-range volume interactions. Therefore, in addition to the potential \eq{eq:01} with the scale-free decay of the coefficients \eq{eq:02}, we add soft repulsive volume interactions, $V({\bf r}_k, {\bf r}_m)$, similar to the ones used in the dissipative particle dynamics \cite{GrootWarren}
\be
V({\bf r}_k, {\bf r}_m) = \begin{cases} c_{int}\,\left( R_{cut} - | {\bf r}_k-{\bf r}_m |\right)^2 & \mbox{for $|{\bf r}_k-{\bf r}_m|<R_{cut}$} \medskip \\
0 & \mbox{for $|{\bf r}_k-{\bf r}_m|>R_{cut}$} \end{cases}
\label{eq:04}
\ee

The choice \eq{eq:04} of a potential for the volume interactions is based on two observations: (i) soft interactions in \eq{eq:04} have the same quadratic form as the fBm potential in \eq{eq:01}, and (ii) the potential \eq{eq:04} allows for some chain self-crossings which speeds-up the chain equilibration and, importantly, produces conformations with the "optimal" amount of entanglements. Following \eq{eq:02}, we set the fBm spring coefficients to $a_{km} = c_{spr}|k - m|^{-\gamma}$, where $c_{spr}$, along with $c_{int}$, is another interaction parameter. Obviously, if $c_{int} \gg 1$ and $c_{spr} \gg 1$, the system in simulations will tend to freeze. To the contrary, if $c_{int} \rightarrow 0$ and $c_{spr} \rightarrow 0$, the system will resemble a disconnected set of beads. In order to be able to analyze knot complexity, we need to make sure that the potential barrier for the chain self-crossing (denoted as $c_{ev}$) satisfies two requirements: (i) $c_{ev}$ is sufficiently high to maintain the overall bead repulsion (at each given point in time, the number of the chain self-crossing events should be small, $n_{cross} \ll N$), and (ii) $c_{ev}$ is low enough to ensure the condition $t_{cross} < T_{sim}$, where $t_{cross}$ is the lifetime of any topological constraint and $T_{sim}$ is the overall time of the simulation trajectory. For the sake of simplicity and because of similarity between the fBm potential and the volume interaction coupling, we set $c = c_{int} = c_{spr} = 3$ as a default for our simulations in the box with periodic boundary conditions (PBC). To probe a robustness of our results to slight variations of the interaction parameters, we have performed auxiliary equilibration runs in the PBC box for a system with a twice higher excluded volume barrier, $c_{int}=6, c_{spr}=3$. Overall, such a choice of parameters has allowed us to simulate a polymer system with excluded volume, but with annealed topology, i.e. topological constraints are subject to equilibrate along with the conformation.

One may wonder how a model system with permitted self-crossings can be used to analyze topology. Considering $\gamma$ (see \eq{eq:02}) as an adjustable parameter, we can select any requested fractal dimension, $\tilde{D}_f$, for the system and equilibrate the chain conformations much faster (due to the presence of self-crossings). In \cite{Schram} a similar approach has been used to equilibrate conformations of long chains in a globular state. When the equilibration is completed, one can obtain an ensemble of conformations with different topological complexities taking the snapshots of the simulation trajectories at some fixed time moment.


Computational complexity of a single simulation for a typical $N$-bead system with long-range interactions \eq{eq:02} grows as $N^2$ for particle-based techniques, such as the molecular dynamics or the Monte-Carlo method. In our simulations we use a slightly modified version of the traditional Metropolis Monte-Carlo (MMC) algorithm \cite{chin} using only local steps assigned for small beads displacements in random directions. Such a version of MMC has been originally developed for the systems with long-range interactions, such as polyelectrolytes, and has been further optimized for computations using graphics processing units (GPU). Its main difference from the traditional MMC used in polymer simulations consists in choosing the beads displacement attempts (from $1$ to $N$ at each step) sequentially, which permits a more efficient parallelization on the GPU hardware.

Using the GPU-optimized algorithm we have been able to simulate $\sim 10^6 \div 10^7$ MCS (Monte Carlo steps, i.e. number of elementary local steps divided by $N$) for the chain length of $N = 16384$ beads. Initial states for MC simulations are prepared as 3D Gaussian random walks without excluded volume. Temperature for the MC step acceptance probability, $P_{acc} = \min\big(1,\exp(-\Delta E/(k_{B}T))\big)$, is set to $k_{B}T = 1$. The cutoff radius of volume interactions equals to $R_{cut}=1$ in all simulations.

Below we consider three setups: (i) chains without excluded volume (fBm polymers), (ii) chains with excluded volume interactions in a free volume, and (iii) ring chains with excluded volume confined in a box with PBC. Since knots are well-defined on closed chains only, rings are better suited for topological analysis. Note that for rings in a PBC box, the long-range harmonic couplings \eq{eq:01} act in the unwrapped space, while volume interactions \eq{eq:04} are computed with the distances to the nearest image at each step. Topological complexity of rings is analyzed using the Alexander polynomials, ${\rm A}(t)$, following the line of reasoning of works \cite{frank, MirnyImakaevNech, Virnau}. To characterize and to compare knotted polymer conformations quantitatively, the so-called "knot complexity", $\chi$, defined as the logarithm of the Alexander polynomial, $\ln {\rm A}(t = -1.1)$, has been used.

\section{Results}

\subsection{Phantom fBm polymers: comparison with the analytical results}

\begin{figure}[ht]
\centering
\includegraphics[width=13cm]{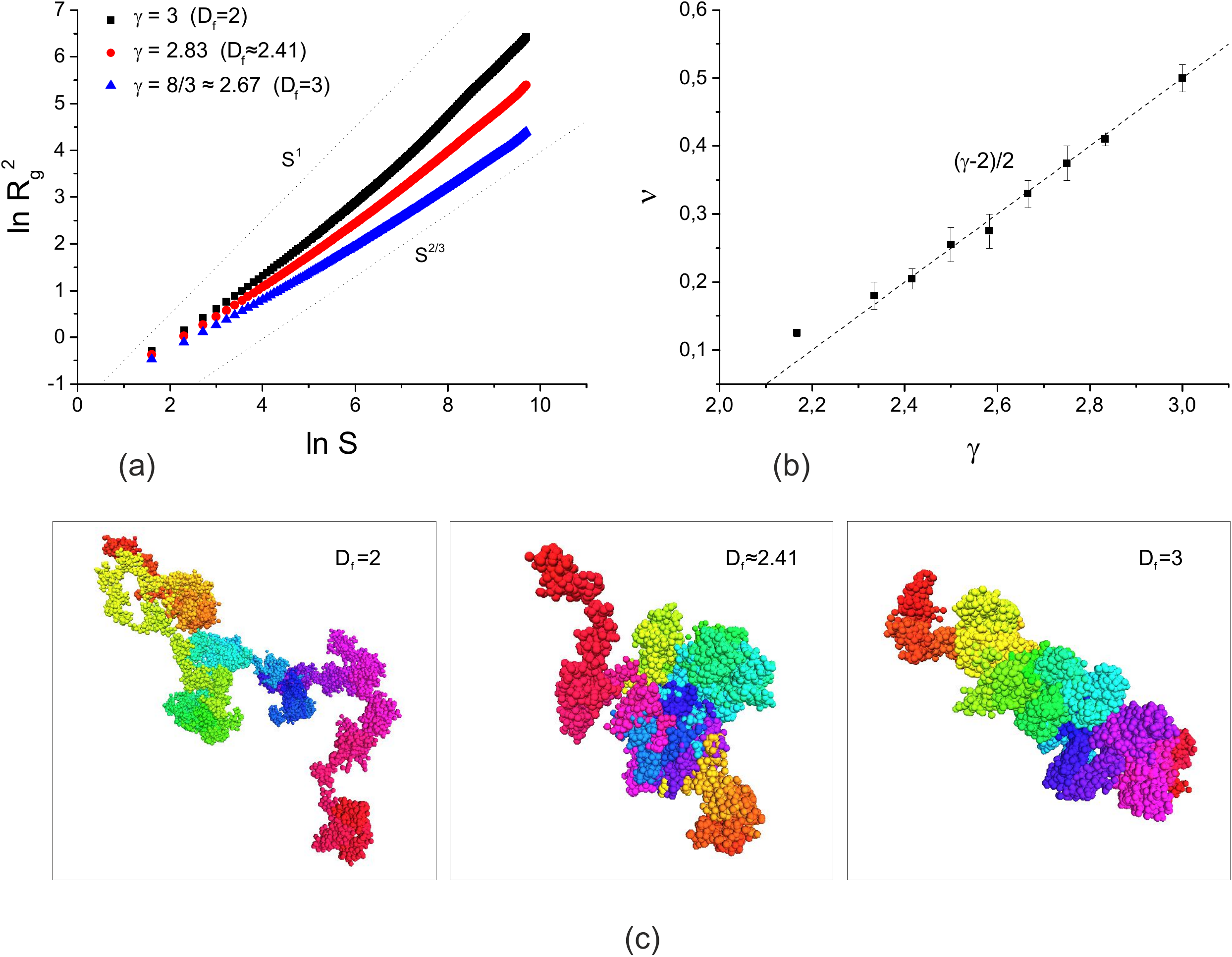}
\caption{Simulations of fBm chains without excluded volume: (a) dependency of the squared gyration radius as a function of the contour distance $R_g^2(s)$ for different $\gamma$; b) Critical exponent $\gamma=1/D_f$ from simulations and the theoretical result \eq{eq:03}; c) Snapshots of equilibrium conformations with different fractal dimension.}
\label{fig:01}
\end{figure}

In order to check our theoretical results on equilibrium statistics of chains equipped with the fBm Hamiltonian \eq{eq:01}, we run Monte-Carlo simulations with quadratic interactions \eq{eq:01} and zero volume interactions, $c_{int}=0$. In our simulations we have varied $\gamma$ controlling the fractal dimension of the polymer chain $D_f$. As it is shown in \fig{fig:01}a,b, simulations perfectly follow our theoretical prediction \eq{eq:03}, i.e the large-scale behavior of chains with the Hamiltonian \eq{eq:01a} is fractal with the fractal dimension $D_f$ \eq{eq:03}. Insignificant deviations from the theoretical line are observed only for very large fractal dimensions $D_f>5$, which do not correspond to any physical system in the three-dimensional space. Such a strong compaction of the chain is associated with large entropic cost, which apparently requires to be balanced by somewhat stiffer harmonic springs (with increased $c_{spr}$ or decreased the effective temperature).

The snapshots of fBm trajectories presented in the \fig{fig:01}c demonstrate their hierarchical folding. At small scales chains resemble blobs of monomers that are brought together by springs, while the large-scale folding is controlled by the value of the exponent $\gamma$. Notably, the chains with the fractal dimension close to $D_f=3$ are conceivably intermingling. This is a direct consequence of the following contact probability scaling $P(s)\sim s^{-1}$ (see Fig. S1). For the fractal chains with the Gaussian distribution of the pairwise distance \eq{eq:01distr} prescribes the following behavior of the contact probability:
\be
P(r_k=r_m) \sim \frac{1}{|k-m|^{3/D_f}}; \quad s=|k-m|
\label{pc_fbm}
\ee
which coincide with mean-field argument \cite{Mirny2009}: the contact probability at distance $s$ is inversely proportional to the volume of the chain segment of the same length. This argument does not hold true in general, however, it is valid for the whole fBm class of trajectories. The relation \eq{pc_fbm} is well reproduced for various values of $\gamma$ in our Monte-Carlo simulations (see Fig. S1).

\subsection{Swollen chains in a free volume}

The effect of excluded volume interactions \eq{eq:04} on statistics of fBm paths \eq{eq:01} can be understood using the following generalization of the standard Flory argument. In order to find the optimal end-to-end distance $R(N)$, we balance the elastic energy of the chain, parameterized by the seed fractal dimension $D_f$, and the volume interactions. Introducing the swelling parameter, $\alpha^2 = \frac{R^2}{b^2 N^{2/D_f}}$, the elastic energy of the fBm chain in the $D$-dimensional space dimension can be written as
\be
F_{el} = \frac{D}{2} \alpha^2
\label{fel}
\ee
which is a direct generalization of the entropy of the ideal chain, provided that the end-to-end distance distribution is Gaussian (note that this is fully consistent with the definition of fBm chains -- see \eq{eq:01distr}). A contribution from the repulsive pairwise volume interactions in $D$ dimensions is
\be
F_{int} \sim B N^2 R^{-D} \sim z \alpha^{-D}
\label{fint}
\ee
where $z$ is a parameter controlling the strength of volume interactions
\be
z \sim B b^{-D} N^{\frac{4-\Theta}{2}}
\ee
and $\Theta = \frac{2D}{D_f}$ is the \emph{generalized dimension} of the fBm chain with Hurst exponent $H=1/D_f$ in the $D$-dimensional space. Balancing \eq{fel} and \eq{fint}, one immediately recovers the critical generalized dimension $\Theta_{cr}=4$, for which volume interactions become negligible and the optimal conformation of the chain is the non-perturbed fBm, i.e. $\alpha = O(1)$. For $D_f=2$ the condition $\frac{2D_{cr}}{D_{f}}= 4$ implies the critical dimension $D_{cr}=4$ of self-avoiding chains. For the fBm with $D_f=3$, the critical space dimension is $D_{cr}=6$. In the subcritical regime, i.e. for $\Theta < 4$, the volume interactions are coupled with the long-range memory imposed by the fBm Hamiltonian, and the critical exponent (Flory exponent) $\nu$ in the relation $R \sim N^{\nu}$ is as follows
\be
\nu = \frac{2\left(D_f+1\right)}{D_f\left(D+2\right)}, \qquad \frac{D}{D_f} \le 2
\label{nu0}
\ee
which recovers the critical exponent of phantom fBm, $\nu \to \frac{1}{D_f}$ as $D_f \to \frac{D}{2}$. In the three-dimensional space from \eq{nu0} we get
\be
\nu_{D=3}\left(D_f\right) = \frac{2}{5}\left(1+\frac{1}{D_f}\right)
\label{nu}
\ee
Equation \eq{nu} yields known limiting cases: a phantom randomly branching polymer with $D_f=4$ regains the ideal statistics when the volume interactions are switched on, $\nu(D = 3, D_f = 4) = \frac{1}{2}$, while for the ordinary Brownian motion with $D_f=2$ one reproduces the Flory critical exponent $\nu=\frac{3}{5}$. Relating the seed fractal dimension, $D_f$, to $\gamma$ \textit{via} \eq{eq:03}, we obtain the expression for the Flory critical exponent, $\nu$, as a function of the parameter $\gamma$ in the fBm Hamiltonian
\be
\nu_{D=3}\left(\gamma\right) = \left\{\begin{array}{cl}
\disp \frac{\gamma}{5} & \quad 2<\gamma\le 3 \medskip \\
\disp \frac{3}{5} & \quad \gamma > 3
\label{eq:flory}
\end{array} \right.
\ee
(Recall that $\gamma$ enters in the dependence $a(s)\sim s^{-\gamma}$ -- see \eq{eq:02}).

\begin{figure}[ht]
\centering
\includegraphics[width=13cm]{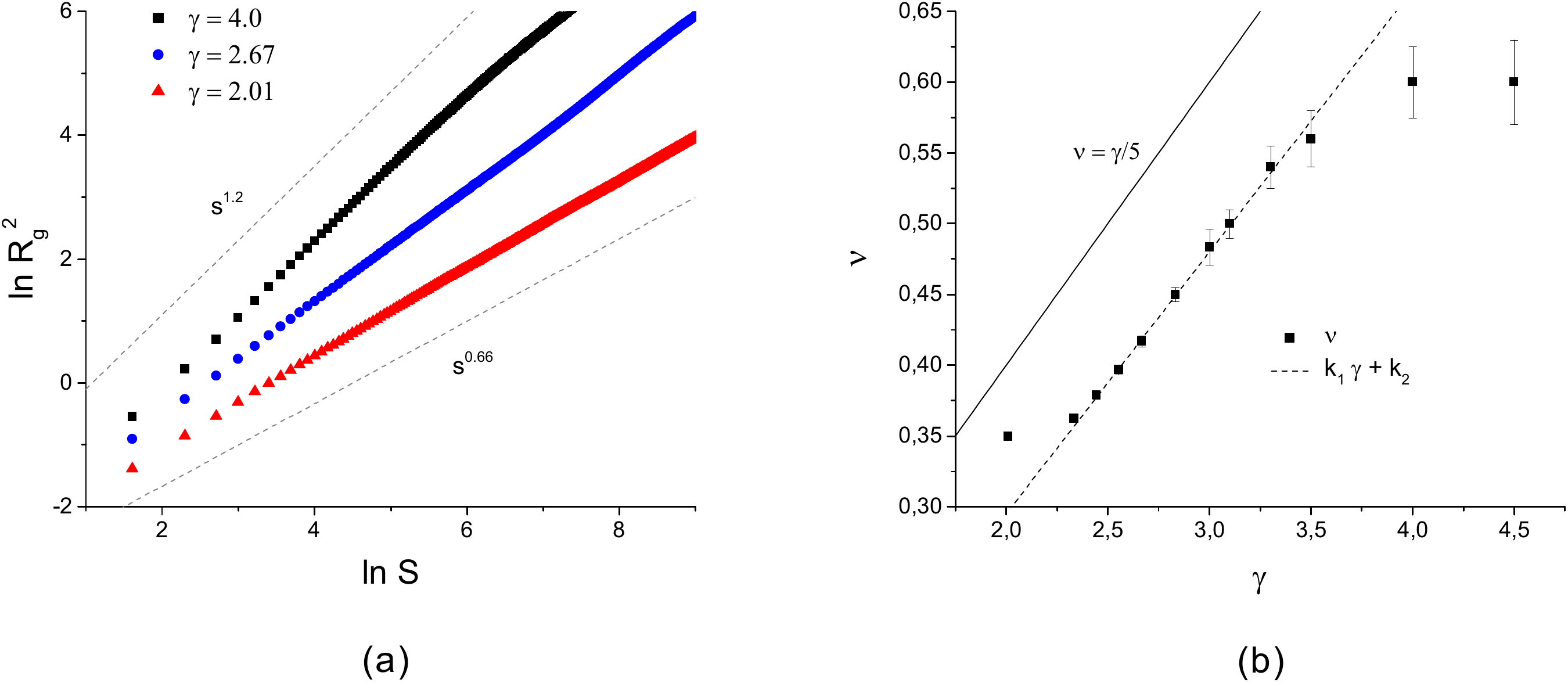}
\caption{Statistics of swollen fBm chains: a) Behavior of the squared gyration radius, $R_g^2$, as a function of the contour distance, $s$, for different $\gamma$; b) Critical exponent $\nu(\gamma)$ from the simulations (points) and the corresponding Flory estimate, $\nu = \gamma/5$, (dashed line); $\nu^{lin} = k_1 \gamma + k_2, \; k_1 = 0.185\pm 0.005,\; k_2=-0.075\pm 0.005$.}
\label{fig:02}
\end{figure}

Having the generalized Flory estimate as a benchmark, we carried out simulations for $\gamma\in [2,5]$. Typical plots of $R_{g}^2(s)$ and $\nu(\gamma)$ and are shown respectively in \fig{fig:02}a and \fig{fig:02}b. Inspecting \ref{fig:02}a, we see that the dependencies $R_{g}^2(s)$ for various $\gamma$ demonstrate a stable fractal behavior, $R_{g}^2 \sim s^{2\nu}$, within several orders of magnitude. Such a stability is important for all forthcoming conclusions in which we characterize the swollen fBm ensembles by $\nu$.


The curve $\nu(\gamma)$ obtained in the computer simulations shown in \fig{fig:02}b lies beneath the theoretical Flory prediction $\nu = \frac{\gamma}{5}$ \cite{flory}. However, in the interval $2.3<\gamma<3.3$, the function $\nu(\gamma)$ is linear and can be approximated as
\be
\nu = k_1 \gamma + k_2; \quad k_1 = 0.185\pm 0.005, \quad k_2=-0.075\pm 0.005
\label{eq:05}
\ee
The slope $k\approx 0.19$ in \eq{eq:05} fully agrees with the Flory prediction \eq{eq:flory}. At small values of $\gamma$ in the interval $2 < \gamma < 2.3$ there is a non-linear regime, due to strong attractive fBm forces dominating over volume interactions. Indeed, in the vicinity of $\gamma=2$, the fractal dimension, $D_f$, of the fBm chain without volume interactions diverges, as it follows from \eq{eq:03}, so that chain non-physically shrinks to the point. In reality, this is, of course, impossible due to the repulsion of higher orders between the monomers (terms higher than two-body in the osmotic pressure expansion) persisting for arbitrary small values of the excluded volume. Higher-order interactions are not taken into account in the Flory theory. As one is moving along the line $\nu(\gamma)$ from large $\gamma$ to $\gamma=2$, the critical exponent exhibits a convex behavior. This non-linearity at small $\gamma$ causes the systematic shift $k_2\approx 0.08$ to the values of the critical exponent $\nu$.

\subsection{Melt of fBm rings: statistics and analyses of knotting}


In order to understand the interplay between knotting and fractal dimension of polymer conformations, we turn to ring chains in a confined volume with a fixed density. Specifically, we consider an ensemble of $N=16384$--bead polymer rings with the stiffness parameter $c_{spr}=3$ and two values of the excluded volume parameter, $c_{ev}=3$ and $c_{ev}=6$. The strength of the volume interactions has been chosen following the considerations outlined in the Model Section. Each ring is confined in a 3D box (simulation cell) of the size $25.2 \times 25.2 \times 25.2$ with periodic boundary conditions, resulting in the volume density of order of the unity.

\begin{figure}[ht]
\centering
\includegraphics[width=13cm]{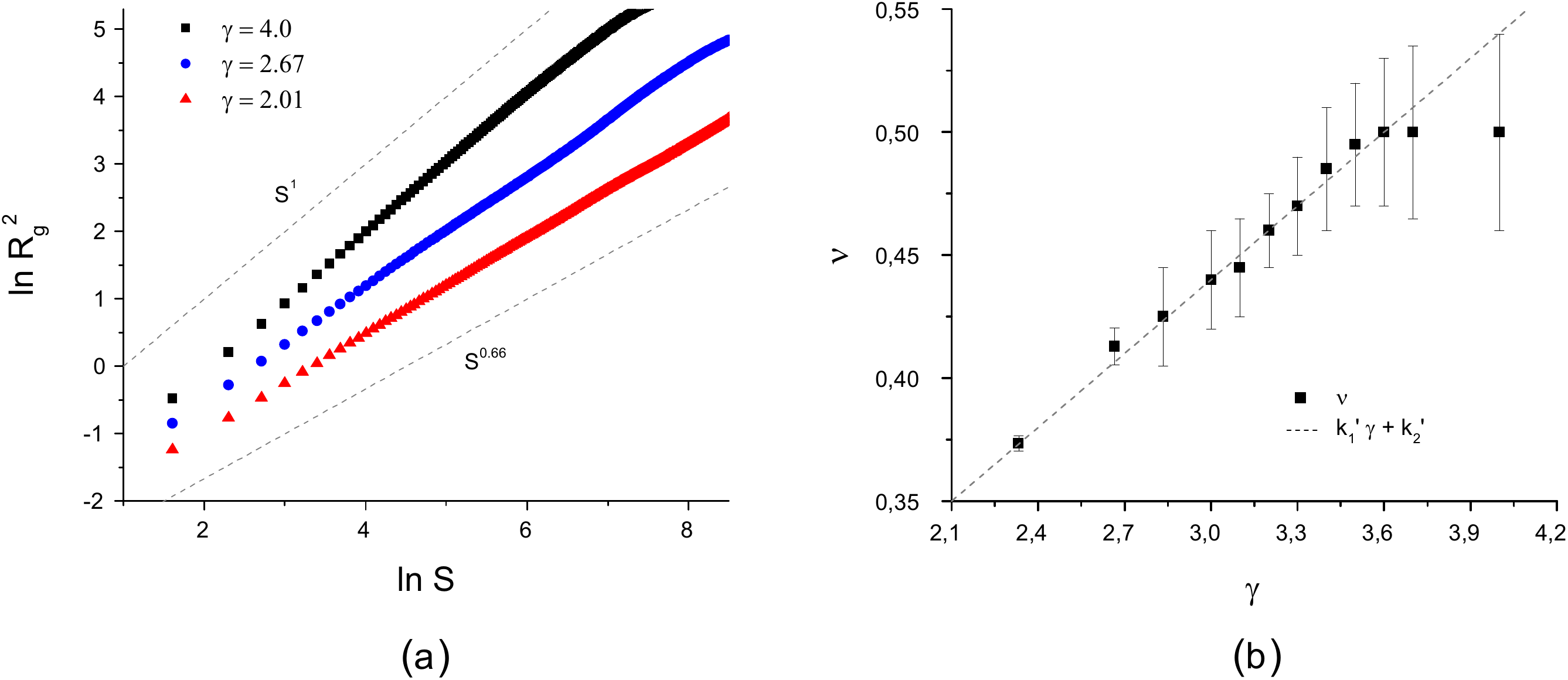}
\caption{Statistical properties of fBm rings with excluded volume in the PBC confinement: (a) plots ${R_{g}}^2(s)$ for different values of $\gamma$; (b) the master curve $\nu(\gamma)$; $\nu(\gamma) = k_1'\gamma + k_2',\; k'=0.10 \pm 0.01,\; k_2'= 0.15 \pm 0.01$.}
\label{fig:03}
\end{figure}

First, in \fig{fig:03}a,b we demonstrate the conformational statistics of rings in the box for $c_{ev}=3$: the squared gyration radius, $R_{g}^2(s)$, and the corresponding critical exponent $\nu(\gamma)$. Simulations points follow again a straight line, as in \fig{fig:01}b, however, the slope is different
\be
\nu(\gamma) = k_1'\gamma + k_2'; \quad k'=0.10 \pm 0.01, \quad k_2'= 0.15 \pm 0.01
\label{eq:06}
\ee
Clearly, the Flory theory is not applicable for chains in the box, since the excluded volume is screened out due to the high monomer density. The logic behind the behavior of the critical exponent $\nu$ here is as follows. In contrast to a linear chain in the free volume, a ring in the PBC box does not swell and thus at $\gamma=2$ (at the point where the ordinary fBm without volume interactions gets collapsed into the point) the ring attains the most possible compact conformation in 3D, i.e. the one with the critical exponent $\nu=1/3$ (and not with $\nu=2/5$ as the Flory theory predicts \eq{nu}). On the other hand, at $\gamma=3 \div 3.5$ the ordinary fBm without excluded volume is ideal and so does a ring with excluded volume interactions in the PBC box. These two points determine the behavior of $\nu(\gamma)$ at all intermediate values of $\gamma$ and explain the observed slope in the simulations, \eq{eq:06} $k' \approx (0.5-0.33)/(3.5-2) \approx 0.11$.

Note that at a twice higher excluded volume strength $c_{int}=6$ the critical exponents stay the same within the error bar (see Fig. S2). That points out to the universality of the critical exponent of fBm to the changes of the excluded volume in simulations with soft volume interactions. We conclude that changing $\gamma$, one could obtain statistically different ensembles of chains described by $\nu$ within the interval $0.35 \lesssim \nu \lesssim 0.5$. Convergence to the equilibrium globule regime ($\nu = 0.5$) is observed for sufficiently slow decaying springs, $\gamma > 3.5$.

\begin{figure}[ht]
\centering
\includegraphics[width=13cm]{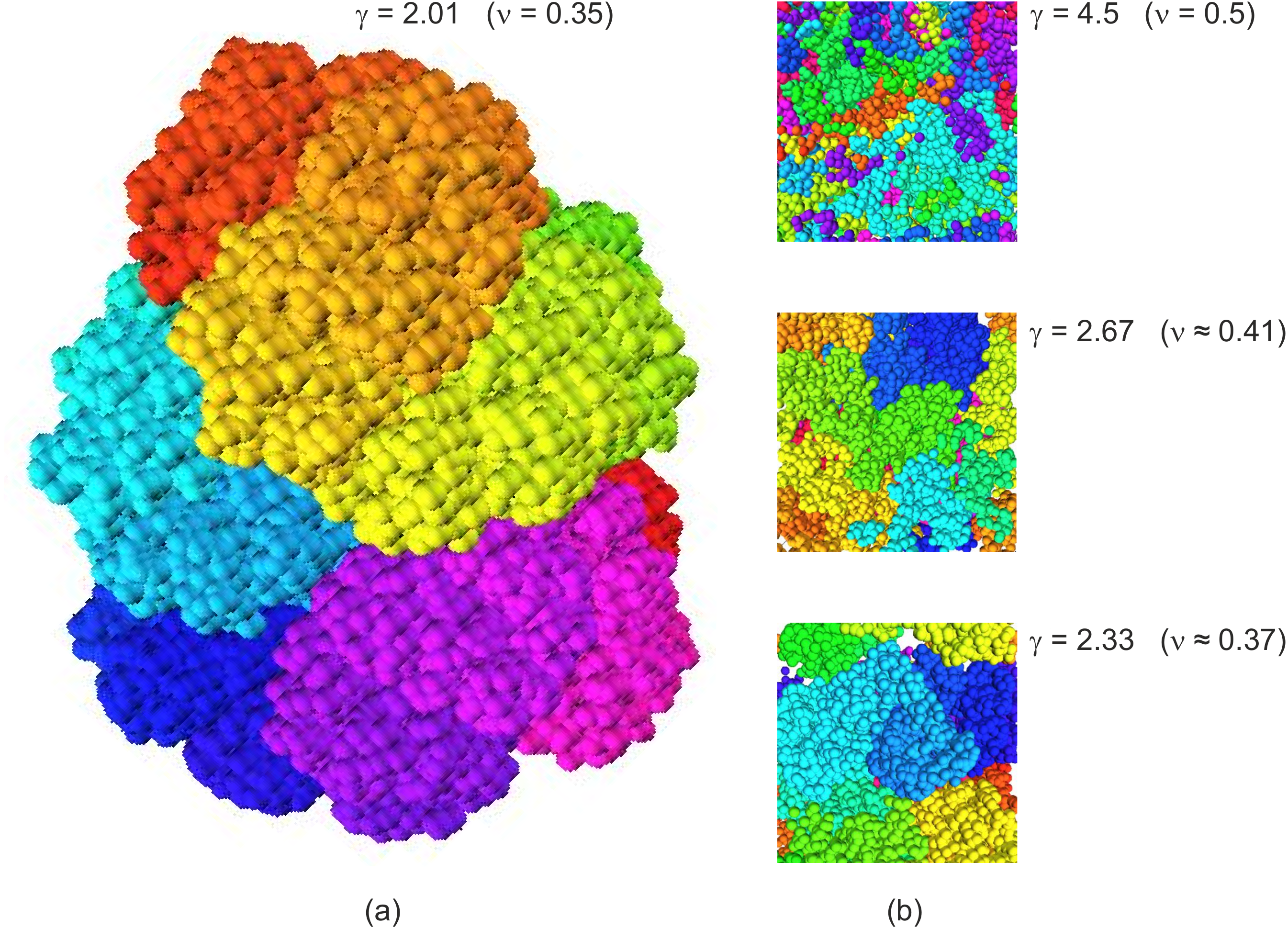}
\caption{Snapshots of fBm chains with excluded volume in the box: (a) $\gamma = 2.01$ corresponding to the fractal dimension $1/\nu \approx 2.86$, (b) $\gamma = 2.33,\,2.67,\,4.5$ with the corresponding critical exponents. Painting chain between extremities in continuously changing RGB colors, we highlight the territorial organization of the path typical for crumpled globule.}
\label{fig:04}
\end{figure}

Snapshots of the conformations in the box are shown in \fig{fig:04}a, where the spatial territorial organization of the chains is clearly seen. The value $\gamma=2.01$ corresponds to the scaling exponent $\nu_e\approx \frac{0.7}{2}=0.35$ and the corresponding effective fractal dimension is $\frac{1}{\nu}\approx 2.86$. Let us note the strong chain "compartmentalization" seen in \fig{fig:04}a typical for the crumpled globule: the nearest neighboring (along the chain) monomers try to be also the nearest neighbors in the space. The snapshots of ring chains in the PBC \fig{fig:02}b indicate the tendency to have more uniform spatial distribution of beads in the box and worsening of the territoriality upon increase of $\gamma$. The corresponding fractal dimensions are obtained from the calibration curve in \fig{fig:03}b.

Despite the fractal dimension $D_f\approx 3$ can be achieved at sufficiently small $\gamma$ for fBm in the box, there is a quantitative difference between rings in a melt and our swollen fBm chains. To demonstrate that we have calculated the contact probability for different values of $\gamma$ both for $c_{ev}=3$ and $c_{ev}=6$ (see Fig. S3). We have observed that upon increase of the fractal dimension, the contact probability exponent, $\alpha$, decreases insignificantly: it varies in the range $\alpha \in [-4/3, -3/2]$ and clearly does not achieve the values relevant for the rings ($\alpha \approx 1 \div 1.1$). Thus, surfaces of crumples in our system are much more smooth than in the ring system with true topological interactions, which is also evident from the snapshots in \fig{fig:04}.

The knot complexity analysis we anticipate by the discussion of the impact of $\gamma$ on the local system density. In Fig. S4 we have collected the distributions of the local density (measured in beads per elementary volume) in chunks of size $\ell_x=\ell_y=\ell_z = \frac{\ell_{box}}{10}$ at different values of $\gamma$. At $\gamma = 2.33$ there is a significant amount of empty chunks (where local density is zero), which means that the typical ring size is essentially smaller than the size of the PBC box. However, for $\gamma \ge 2.67$ the amount of empty chucks vanishes suggesting that the PBC box is completely filled by chain monomers and the density of ring monomers in the PBC box is uniform.

\begin{figure}[ht]
\centering
\includegraphics[width=13cm]{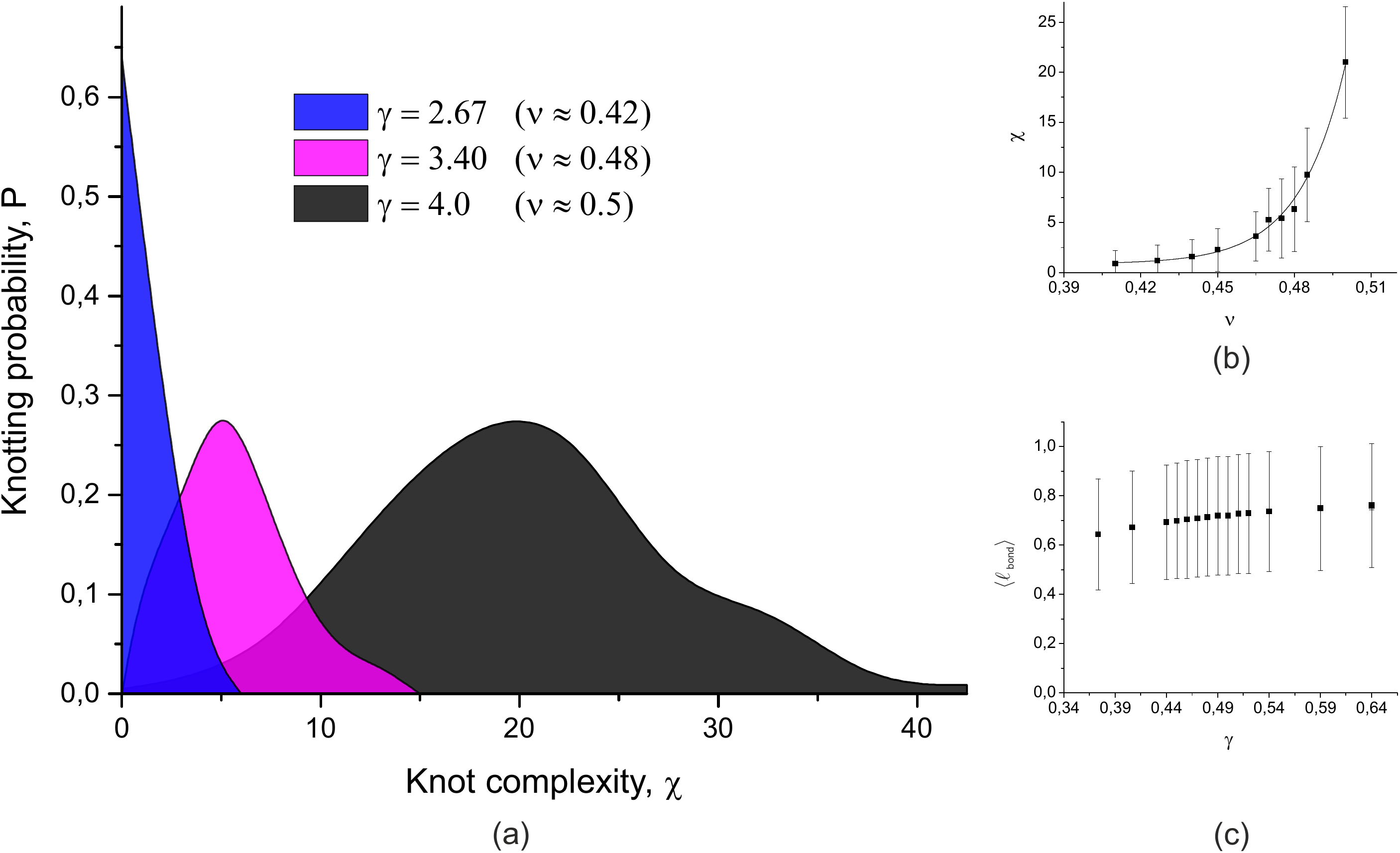}
\caption{(a): Distribution of the knot complexity $\chi$ for fBm rings of N=16384 monomers in PBC confinement with $\gamma=2.67$ (blue), $\gamma=3.4$ (pink) and $\gamma = 4$ (black); (b) Plot $\chi(\nu)$ for fBm rings in the simulation cell with periodic boundary conditions along with the exponential fit; (c) The average bond length $l_{bond}$ as a function of $\gamma$.}
\label{fig:05}
\end{figure}

Now we proceed to the topological analysis of fBm rings with volume interactions. A topological state of a loop can be characterized by a widely used "knot complexity" $\chi = \ln \mathrm{A}(t)$, defined as the logarithm of the Alexander polynomial, $\mathrm{A}(t)$, evaluated at $t=-1.1$ (see, for example, \cite{frank,Virnau}). Sample distributions of knot complexities of loops averaged over single trajectories for $c_{ev}=3$ are depicted for three different values $\gamma=\{2.67, \,3.4,\, 4.0\}$ in \fig{fig:05}. According to the calibration curve in \fig{fig:03}, $\gamma=2.67$ yields conformations with $\nu \approx 0.41$, which is not far away from the critical exponent of the crumpled globule, $\nu=0.33$, while $\gamma=4$ produces equilibrium globule with $\nu \approx 0.5$. Distributions in \fig{fig:05} clearly demonstrate that the knot complexity fluctuates due to the chain self-crossings, however the difference between the distributions for almost crumpled (peaked, blue) and the Lifshitz (wide, black) globules is notable. As discussed above, at values $\gamma < 2.66$ chains do not occupy fully the volume of the periodic box. Thus, to exclude the (possible) nonphysical impact of the occupied volume on the chain self-knotting, we exclude values $\gamma<2.66$ from our knotting analysis. For the stronger excluded volume $c_{ev}=6$ the distributions of the knot complexity similarly depend on the resulting fractal dimension of the chain (see Fig. S5). A found difference between the distributions at $c_{ev}=3$ and $c_{ev}=6$ for $\gamma=2.67$ (Fig. S5(a)) is due to the fact that a chain is able to more easily cross itself at $c_{ev}=3$, being in a dense almost compact state, which ends up in a more fluctuating chain topology. However, this difference does not alter the general tendency of the knot complexity distributions at varying fractal dimensions.

A striking difference in topological properties of globally knotted and unknotted chains is widely known and can be derived from the general statistical behavior of so-called matrix–valued Brownian Bridges (BB) -- see \cite{MirnyImakaevNech}. The knot complexity $\chi$ of loops in the topologically unconstrained globule is expected to grow as $\chi(N) \sim N^2$. Contrarily, due to the global topological constraint imposed in the unknotted globule, its knot complexity grows essentially slower, as $\chi(N) \sim N$, which follows from the statistical behavior of BB in spaces of constant negative curvature. To get an estimate $\chi(\nu)$ for an arbitrary critical exponent $1/3 < \nu < 1/2$, we assume that the complexity $\chi$ obeys a scaling behavior $\chi(\nu) \sim N^{f(\nu)}$ and the simplest form of the function $f(\nu)$ in the range of interest is linear. Having values of $f$ at two limiting points $\nu=1/2$ and $\nu=1/3$ permits us to propose the following scaling relation for the knot complexity
\be
\chi \propto \exp(u\, \nu), \quad u=6\log N\approx 58
\label{chi_sc}
\ee

To analyze the dependence $\chi(\nu)$ in simulations, we have collected snapshots from a number of independent runs. Resulting values of $\chi$, computed for the collected sets of conformations, have been averaged for each value of $\gamma$ (i.e. for the corresponding $\nu$ from \fig{fig:03}). The resulting plot, $\chi(\nu)$, is shown in \fig{fig:05}b. Note that at $\nu < 0.45$ the knot complexity $\chi$ vanishes, which indicates that the chain conformations are weakly knotted. At $\nu > 0.45$ we observe a rapid exponential growth of $\chi$. The exponential fitting of $\chi(\nu)$ obtained from our simulations provides
\be
\chi = \chi_{0} + A \exp(u\, \nu), \qquad A = (1.5 \pm 0.2 )\times 10^{-11}, \quad \chi_{0}=0.9 \pm 0.2, \quad u = 56 \pm 3
\ee
The value of $u$ is in the perfect agreement with the scaling estimates \eq{chi_sc}.

Some other factors apart from $\nu$, e.g. the change of the average bond length, might influence the knot complexity in our simulations. From \cite{MirnyImakaevNech} it is known that in the equilibrium globule $\chi \sim L^2$, while in the crumpled one $\chi \sim L$, where $L=Nl_{bond}$ is the chain length, $N$ is the number of bonds and $l_{bond}$ is the average bond length. In the fBm with different fractal dimension the consecutive beads are linked by springs with different rigidities, thus, the parameter $l_{bond}$ is also $\gamma$-dependent. However, as it is shown in \fig{fig:05}c, $\ell_{bond}$ grows only sub-linearly with $\gamma$. Therefore, we conclude that the bond stretching has a negligible effect on the chain self-knotting.

\section{Discussion}

In our work we have developed a machinery permitting to operate with fractal chains with volume interactions and producing compact chain conformations with the fractal dimension $D_f\ge 2$ in the three-dimensional space. Our study is motivated by an attempt to mimic the conformational statistics of collapsed nonconcatenated polymer loops which are known to form the compact hierarchical crumpled globules (CG) with the fractal dimension $D_f=3$ at large scales. The CG structure manifests itself in many applications, including the "territorial organization" of human chromatin in the cell nucleus, however due to the non-local topological interactions, analytic and numeric treatment of CG is an extremely difficult problem. In our work we have addressed the following question: "Is it possible to reproduce the topologically-stabilized CG by a self-avoiding fBm chains with a properly adjusted fractal dimension?" If the self-avoiding fBm paths are able to reproduce the basic features of collapsed unknotted polymer rings, we would tremendously simplify the generation of CG-like conformations since the topological constraints could be completely washed out from the consideration.

Our study largely stems from the work \cite{fbm1}, where it was shown that fractional Brownian motion with various Hurst exponents could be constructed using a hierarchical pairwise quadratic potential \eq{eq:01} with the spring constants $c_{spr} \sim |k-m|^{-\gamma}$. In the current work we verify our main theoretical result, a dependency of the Hurst exponent of the paths on the value of $\gamma$, in the Monte-Carlo simulations.

A special attention we paid to the correct account of volume interactions added on top of fBm for both open chains in a free space, and ring polymers confined in the box with the periodic boundary conditions. In presence of soft volume interactions, the value of $\gamma$ controls the value of the critical exponent of the resulting equilibrium conformations, which ranges from crumpled globules ($\nu \approx 1/3$) to either swollen coils (open chains in the free volume, $\nu \approx 0.58$) or equilibrium globules (ring chains confined in the PBC, $\nu=1/2$). We have calibrated the model and found that both cases, linear chains in the free volume, and ring chains in the PBC exhibit a linear relationship between $\gamma$ and the critical exponent $\nu$. We provide the generalized Flory-type arguments for the linear chains in the free volume, which are well consistent with our Monte-Carlo simulations.

Using the developed technique we have analyzed the relation between the fractal dimension, $D_f$, of fBm rings with volume interactions in the box and their knot complexity $\chi$. We have shown that with increase of $D_f$, the typical chain conformations become less and less knotted. Besides, we have established the dependence of $\chi$ on the Flory critical exponent, $\nu$, both from simulations and using scaling arguments. Overall, we have shown that using the self-avoiding fBm model without any direct account of topological interactions, one can generate and study weakly knotted compact conformations with $2\le D_f<3$.

Though the fractal dimension of chains with fBm-like memory and volume interactions in the confining box can attain values close to $D_f \approx 3$, typical for unknotted rings in the melt, the former seem to be qualitatively different from the latter. The "compartmentalization" depicted in \fig{fig:04}a implies territorial organization of the crumples in the sense of the fractal dimension, however, the corresponding boundaries of the territories are found to be very smooth. By a direct construction, the work \cite{Smrek13} has provided a whole family of statistically different paths, all having the same fractal dimension $D_f=3$. The physical difference between these paths consists in a different level of intermingling between the crumples. Rings have sufficiently developed surfaces of crumples, while the Hilbert (Peano) curve has smooth surfaces. This property of the surface smoothness is reflected in the behaviour of the contact probability $P(s) \sim s^{-\alpha}$ with $\alpha\approx 1$ for rings and $\alpha=4/3$ for the Hilbert curve. For fBm rings the values of the contact probability exponent vary (with the change of the induced fractal dimension) rather subtly between 4/3 and 3/2, implying sufficient smoothness of boundaries between distinct crumples. We refer this observation to the fact that topological interactions are essentially collective, i.e. there are effective interactions between three, four etc monomers. The fBm Hamiltonian constructs fractal paths by means of quadratic interactions only, which along with the pairwise excluded volume interactions optimize the chain conformations locally. However, the approach presented in this paper allows to mimic a similar "territorial" organization in sense of the fractal dimension without any real topological constraints.

One of clear computational disadvantages of our model is that, from the computational standpoint, we deal with an $N$--body problem equipped with long-range interactions, hence the possible size of the systems is fairly limited. It might be beneficial to explore different simulation techniques, such as molecular dynamics, which is inherently better suited for parallelization of the computations.

\begin{acknowledgements}
We are grateful to A. Grosberg, L. Mirny, R. Everaers, A. Rosa and M. Tamm for fruitful discussions. The work of AMA and VAA was partially maintained within frameworks of the state task for the FRC CP RAS \#FFZE-2019-0016; SKN and KEP acknowledge the support of the grant RFBR 18-29-13013.
\end{acknowledgements}

\begin{appendix}

\section{Supplement}

\begin{figure}[ht]
\centering
\includegraphics[width=10cm]{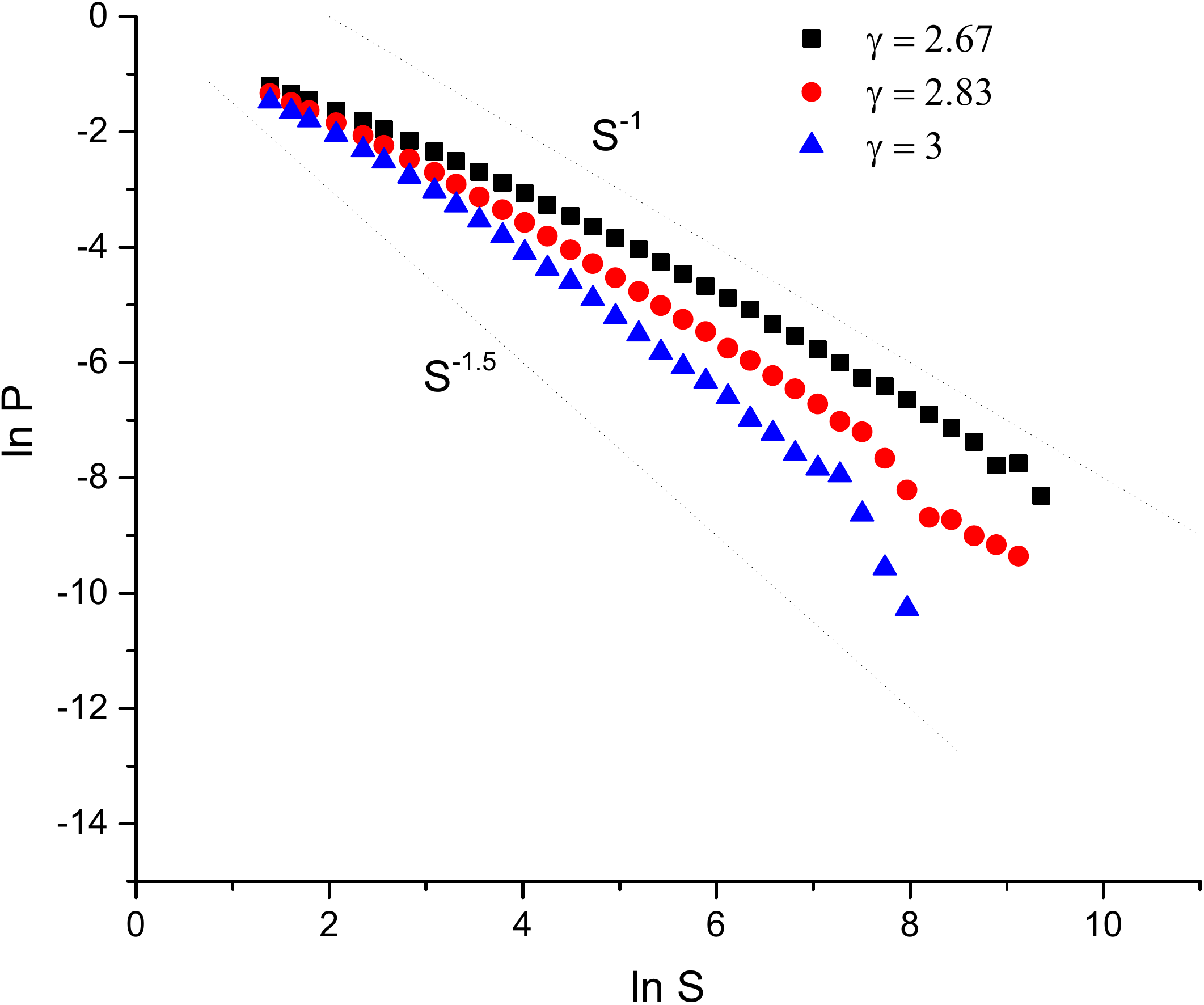}
\caption{}
\end{figure}

\begin{figure}[ht]
\centering
\includegraphics[width=10cm]{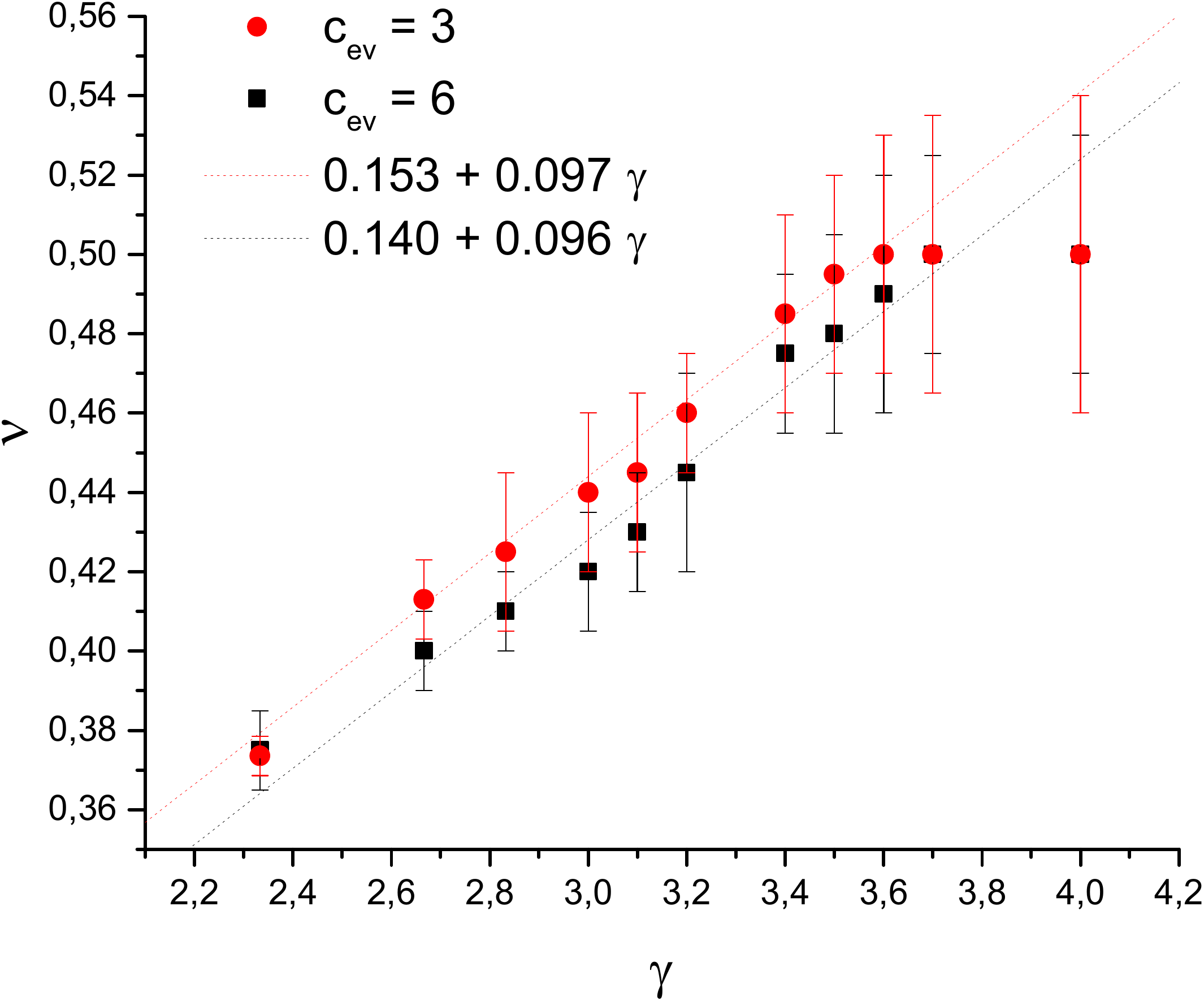}
\caption{}
\end{figure}

\begin{figure}[ht]
\centering
\includegraphics[width=13cm]{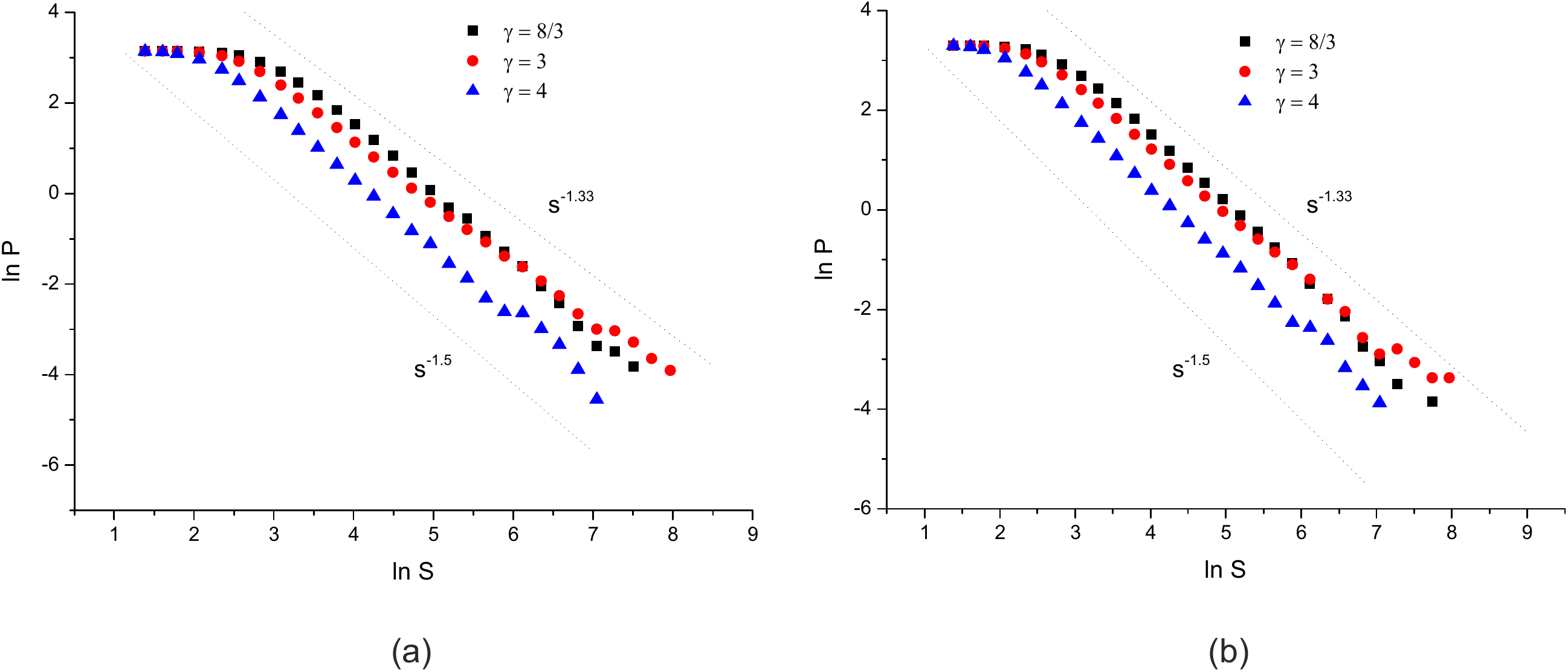}
\caption{}
\end{figure}

\begin{figure}[ht]
\centering
\includegraphics[width=10cm]{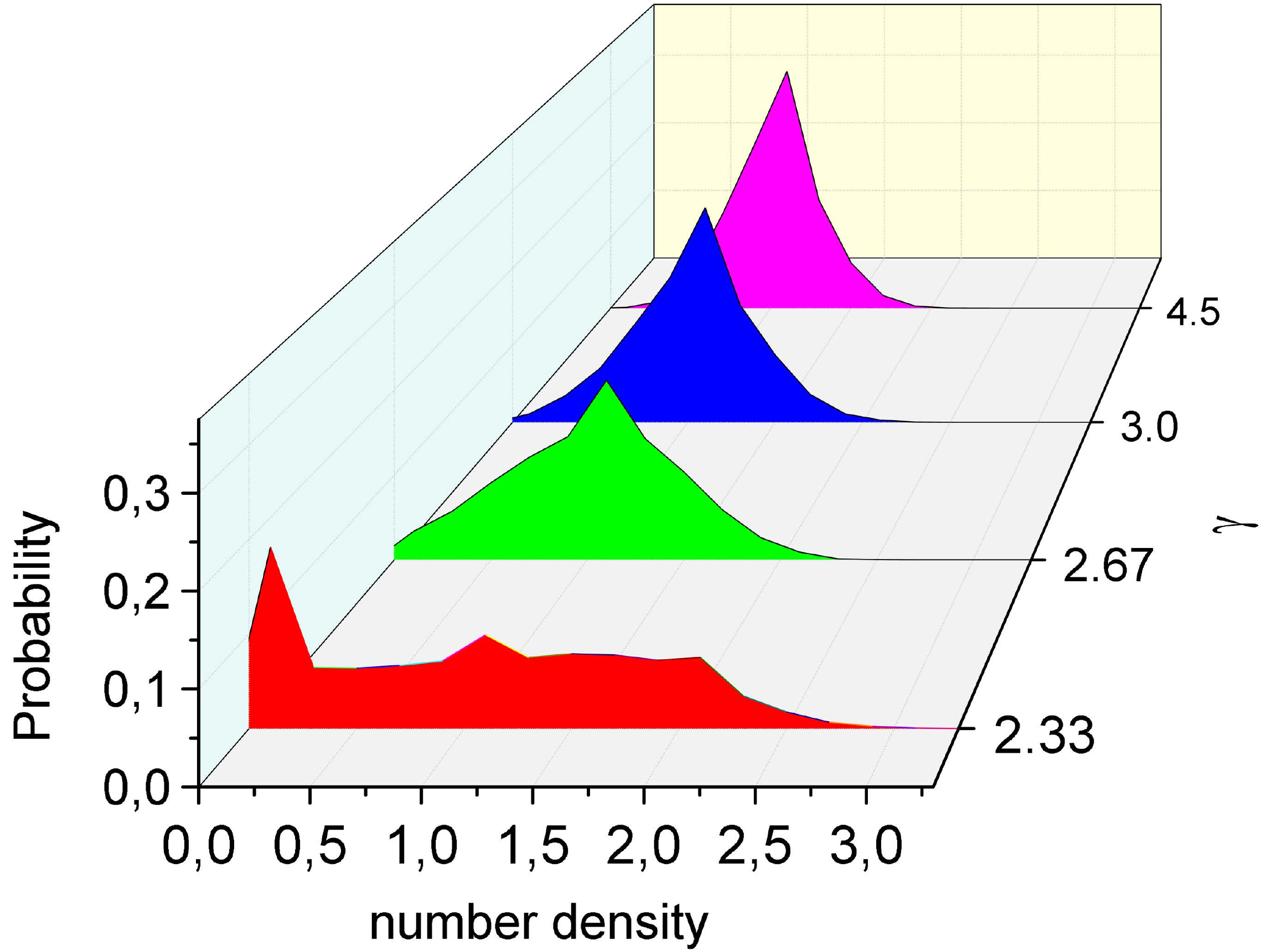}
\caption{}
\end{figure}

\begin{figure}[ht]
\centering
\includegraphics[width=13cm]{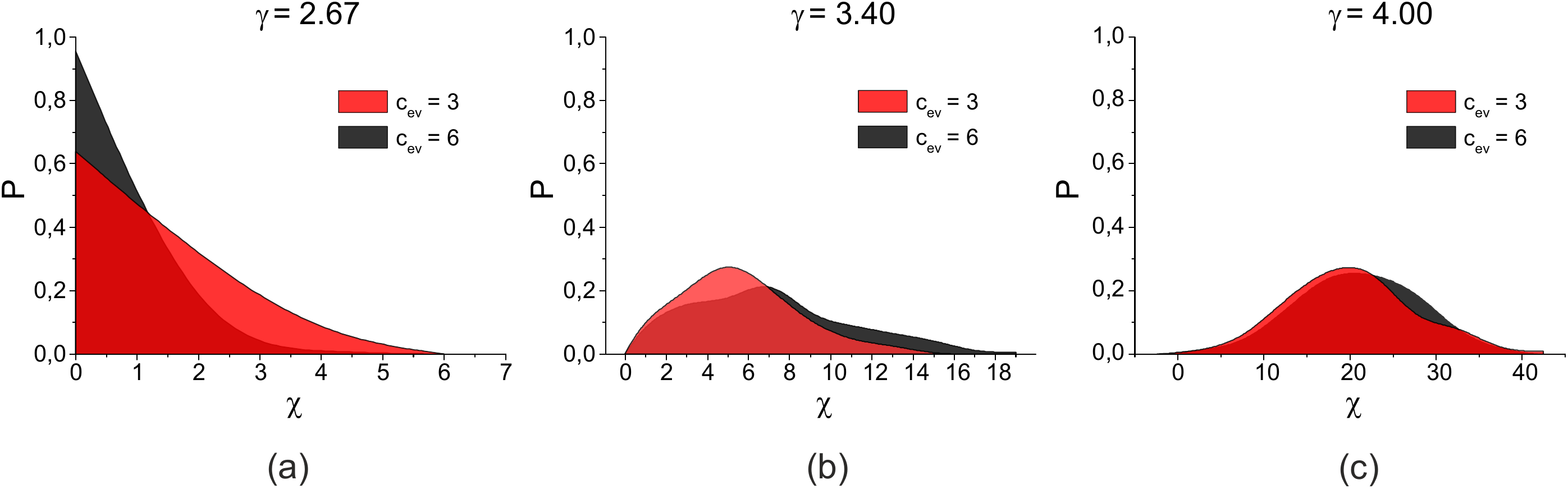}
\caption{}
\end{figure}

\end{appendix}

\end{document}